\documentclass[showpacs,pre,twocolumn,floatfix]{revtex4}

\usepackage{graphicx}
\usepackage{dcolumn}
\usepackage{bm}

\usepackage{amssymb}
\usepackage{amsmath}
\begin{document}


\title{Reply to Comment on ``Existence of Internal Modes of sine-Gordon Kinks''}

\author{Niurka R. Quintero} \email{niurka@euler.us.es}
\affiliation{Departamento de F\'\i sica Aplicada I, E.U.P.,
Universidad de Sevilla, c/ Virgen de \'Africa 7, 41011 Sevilla,
Spain}

\author{Angel S\'anchez}%
\homepage{http://gisc.uc3m.es/~anxo}
\affiliation{%
Grupo Interdisciplinar de Sistemas Complejos (GISC) and
Departamento de Matem\'aticas, Universidad Carlos III de Madrid,
Avenida de la Universidad 30, 28911
Legan\'es, Madrid, Spain, and\\
Instituto de Biocomputaci\'on y  F\'{\i}sica de
Sistemas Complejos (BIFI),\\
Facultad de Ciencias, Universidad de Zaragoza, 50009 Zaragoza,
Spain}

\author{Franz G.\ Mertens}
\email{Franz.Mertens@uni-bayreuth.de} \affiliation{Physikalisches
Institut, Universit\"at Bayreuth, D-85440 Bayreuth, Germany}

\date{\today}

\pacs{05.45.Yv, 03.50.-z}

\begin{abstract}
In this reply to the comment by C.\ R.\ Willis, we show, by
quoting his own statements, that the simulations reported in his
original work with Boesch [Phys.\ Rev.\ B {\bf 42}, 2290 (1990)]
were done for kinks with nonzero initial velocity, in contrast to
what Willis claims in his comment. We further show that his
alleged proof, which assumes among other approximations that kinks
are initially at rest, is not rigorous but an approximation.
Moreover, there are other serious misconceptions which we discuss
in our reply. As a consequence, our result that quasimodes do not
exist in the sG equation [Phys.\ Rev.\ E {\bf 62}, R60 (2000)]
remains true.
\end{abstract}

\maketitle

\section{Introduction}

In his comment \cite{williscom}, Willis claims that in \cite{bw}
he used a ``rigorous'' projection operator collective variable
formalism for nonlinear Klein-Gordon equations to ``prove'' the
continuum sine-Gordon (sG) equation has a long-lived quasimode
whose frequency $\omega_s= 1.004$ is in the continuum just above
the lower phonon band edge with a lifetime $(1/\tau_s) = 0.003
\Gamma_0$. In \cite{us1} we performed two numerical investigations
which show that neither intrinsic internal modes nor quasimodes
exist in contrast to previous results. Willis further claims that
he ``proves'' that our first numerical investigation could not
possibly observe the quasimode in principle and our second
numerical investigation actually demonstrates the existence of the
sG quasimode. He states that his analytic calculations and
verifying simulations were performed for a stationary sine-Gordon
soliton at the origin, but that we explained our analytic
calculations and confirming simulations in terms of the Doppler
shift of the phonons when stationary sine-Gordon solitons have a
zero Doppler shift. In the following we argue that his comment is
irrelevant because the results Willis refers to in \cite{bw} were
obtained for {\em moving kinks} (which we prove by quoting
statements from \cite{bw}), and because his alleged ``proof'' is
not a mathematical proof but an approximate calculation as any
collective coordinate theory \cite{SIAM}.

\section{On the non-stationarity of kinks}

After carefully reading their paper \cite{bw} several times, we
are afraid that Willis's comment suffers from severe
misconceptions and, what is worse, that the author is not familiar
enough not only with our work \cite{us1} but also with his own
paper \cite{bw}. Quoting literally from \cite{bw}, p.\ 2296, first
paragraph beginning after Eq.\ (3.2), it reads: ``We impose
initial conditions in our simulations in two different ways. The
first way is by specifying the field [\dots] thus giving the
equilibrium kink a nonzero initial velocity". And subsequently,
three out of four figures of the paper refer to this type of
simulation. For instance, the caption of Fig.\ 1 reads: ``The
initial condition for $t=0$ is a kink with velocity determined by
specifying the kink shape at\dots". Therefore, in spite of
Willis's claims that their simulations were for stationary kinks,
that is not the case, and he cannot dismiss our explanation of his
observations based on this. Kinks moved in his simulations and
correspondingly their phonon spectra were shifted, as we explained
in \cite{us1}.

\section{On Willis's misconceptions}

Given the fact that the author is not aware of his own work and
results, it is no surprise that he is completely unfamiliar with
\cite{us1}. His comment \cite{williscom} is full with wrong
statements about what we reported in that paper. Thus, he claims:

\begin{itemize}
\item ``What they measured was the phonon absorption spectrum of
the linearized sine-Gordon (sG) equation\dots" This is simply not
correct. We simulated the full sG equation: in page R61 of
\cite{us1}, in the beginning of section III, this is perfectly
clear: ``We have computed the numerical solution of the perturbed
sG equation...". We never linearized anything and worked with the
full equation.

\item ``Consequently, their two equations of motion for P(t) and
l(t) which they state is the `basis of their theoretical analysis'
of Ref.[3] have absolutely no relevance to our analytic derivation
and confirming simulations." This is not correct. Actually, those
equations are the way we find that, were there any internal modes
or quasimodes in the kink, they would be excited parametrically by
the ac driving. In spite of his claim that he ``proved in Sec.III
that an ac driver can not create a sG quasimode'', we have shown,
by means of those equations, that parametric excitation at half
the frequency of the mode must be observed. We refer the
interested reader to \cite{us2,us3,us4} to learn more about this
effect. On this issue, it is particularly important to realize the
inconsistencies in Willis's reasoning: Just after Eq.\ (16) of
\cite{williscom} Willis claims: "Thus an ac force field cannot
possibly excite a phonon mode. Consequently, their first numerical
investigation in ref [3] could not possibly detect the presence of
the sG quasimode and thus it has not relevance whatsoever to the
existence or nonexistence of the sG quasimode". If this were true
(which is not), Willis would be forced to conclude that a
biharmonic force $f(t)=-\left[\epsilon_1 \cos (\omega t)+
\epsilon_2 \cos( 2 \omega t+\theta)\right]$ is also unable to
excite the phonons. But this is not the case: On the contrary,
Willis has recently studied \cite{pepe}, by using the same
approximate method that he developed in 1990, the action of this
biharmonic force and dissipation on the ratchet motion of the
kink. In this paper, he claims: ``We use a rigorous collective
variable for nonlinear Klein-Gordon equations to prove that the
rectification of the current is due to the excitation of an
internal mode $\Gamma(t)$, which describes the oscillation of the
slope of the kink, and due to a dressing of the bare kink by the
ac driver.'' In Willis's own words, dressing of the kinks means
phonons excited in the system, in blatant contradiction with his
remark on our work. We also note that Willis's discussion of our
theoretical treatment is plagued with his misconception about
``exact analytical calculations'' in his earlier paper, which
unfortunately are not exact (cf.\ Sec.\ V below).

\item ``The simulations and analysis by the authors in Ref. [3]
were done for an appreciably discrete sG equation". This is not
correct. Our discretization was of the same order as that of
\cite{bw}: 0.05 in our case, 0.02 in theirs [cf.\ \cite{bw} p.\
2296, below Eq. (3.1)]. If ours is irrelevant, so is the one in
\cite{bw}, and vice-versa. The author either did not remember the
discretization they used, or has not read our paper carefully
enough to notice ours.

\item On the other hand, Willis does not trust our numerical
simulations and explicitly refers to the ``incompetent design of
their [ours in \cite{us1}] simulation", claims that all we observe
in Fig.\ 2 of \cite{us1} is ``a complicated interference pattern",
arising from ``reflection of phonons at the boundary at time 200".
We were fully aware of the possibility of phonon reflection at the
boundaries, this being the reason why we used free boundary
conditions, to minimize this effect. Furthermore, we did not trust
a visual inspection of our Fig.\ 2 as he does, and carried out a
Fourier analysis of the time evolution of the kink width, finding
only phonons (as Willis points out, present in the system due to
emission and reflection). We did not find any Fourier peak arising
from a hypothetical ``quasimode", in spite of the fact that the
author of the comment concedes that ``the first 200 seconds
[actually, time units] of their simulation of the width l(t)
[actually, the simulation was for the full sine-Gordon equation,
not of the width] gives a very good representation of the sG
quasimode". How can then be that it does not show up in the
Fourier spectra, which we report in the paragraph below Fig. 2?
\end{itemize}

\section{On further evidence on the non-existence of quasimodes}

Having discussed the lack of familiarity of the author of the
comment with both his own work and the one he comments upon, we
would like to refer the reader to research Willis is not aware of,
and that has significantly advanced the subject. Indeed, one of us
(NRQ) has published a very relevant investigation on this subject,
\cite{niurka}, showing that the spectrum mentioned in the comment,
see Eq. (15), and which we used in our own paper [Eq. (15) of
\cite{us1}] is not correct, and in turn the parenthesis should
read $(n-1)\pi/L$ for small $n/L$. This paper is very pertinent to
this discussion because the author of the comment seems to be very
worried that we are not aware of the ``phonon dressing" of the
kink. This work of Quintero and Kevrekidis is the best way to
understand what are the consequences of the phonons on the kink
behavior.

Furthermore, there was only one other paper in the literature that
ever observed something similar to a ``quasimode" \cite{eva}, in
fact using simulations of kinks driven by constant forces. This
only independent confirmation of the ``quasimode" was also proven
wrong by us in \cite{us5}. In this case, it is not a matter of
simulations and how they measured the ``quasimode": In the above
reference we showed that the analytical calculations in this other
paper were not correct. We encourage the reader to take this
further report on the absence of quasimodes into account.

\section{On the approximate character of collective coordinate
theories}

Finally, Willis does not seem to be informed about the nature and
rigor of collective coordinate calculations. He states in the
abstract that ``... prove the continuum sine-Gordon equation has a
long lived quasimode...". Unfortunately and contrary to his
belief, collective coordinate calculations can not prove anything
in the mathematical sense, in so far as they are approximate. Even
if he tries to keep his calculation (which, by the way, he
reproduces literally from \cite{bw}, therefore adding no new
evidence at all in favor of his claims) exact, he cannot, and he
has to make a number of approximations including two
linearizations and assuming $X$ and its derivative to be exactly
zero (see p.\ 2293 of \cite{bw}, 2nd paragraph on the right).
Willis goes as far as claiming: ``their Eq.\ (3) for $l(t)$ is
incorrect because it contains none of the many terms proportional
to $\chi(t)$ that appear in the exact equation of motion\dots''
Our equation is completely correct, consistent with our
derivation, and confirmed by numerical simulations. Our result is
different of his simply because neither of them is exact. They
begin with different {\em Ans\"atze} and subsequently lead to
different {\em approximate} equations. Collective coordinate and
related calculations are interesting, often times very accurate
and useful, but by no means exact and, mathematically speaking,
constitute no proof. Indeed, it has been shown that such a
procedure may lead to wrong predictions: For a detailed
discussion, see \cite{SIAM}.

In this respect, we want to stress that when Willis uses the paper
\cite{kal} in favor of his argument, it is again an approximate
result and a conjecture. It is true that K\"albermann calculates
some solutions analytically, but he has to resort to numerical
simulations to make his point. In addition, the relationship of
K\"albermann's ``wobble solution" with Willis's ``quasimode" is
dubious at best: K\"albermann finds a continuum of such possible
quasimodes, which is not surprising because his wobble solution is
a combination of a kink and a breather, and therefore is a
one-parameter family of solutions; a longer discussion of this
work is out of the scope of this paper, but we want to stress that
he never finds a specific quasimode as Willis claims, and
K\"albermann's only example has a frequency below the phonon band,
very different from Willis's quasimode. Furthermore, this wobble
is, in K\"albermann's own words, only ``apparently stable" and all
he can be positive about is that ``the kink appears to be decaying
to a wobble". Therefore, there is no such thing as a mathematical
proof of the existence of ``quasimodes" in this paper either.

\section{Conclusion}

Based on all the arguments summarized above, we are afraid that
the preceding comment \cite{williscom} is a combination of wrong
memories and misconceptions about our work. Willis is not aware
that his own work did not deal with stationary kinks, he is not
familiar with the fact that the simulations in \cite{bw} are as
discrete as ours, does not understand that collective coordinate
theories are not rigorous mathematical proofs, is not aware of
further developments on this question, and reads our paper
\cite{us1} in an unscientifically biased manner. Therefore,
Willis's comment does not change our conclusions, and our result
that quasimodes do not exist in the sG equation \cite{us1} stands
in its own right.

 \subsection*{Acknowledgments}
This work has been supported by the Ministerio de Ciencia y
Tecnolog\'\i a of Spain through grants FIS2005-973 (NRQ),
BFM2003-07749-C05-01, FIS2004-01001 and NAN2004-09087-C03-03 (AS)
and by ``Acciones Integradas Hispano-Alemanas''
HA2004-0034---D/04/39957".

\end{document}